\documentclass[aps,prl,longbibliography,superscriptaddress,twocolumn,amsmath,amssymb,showpacs,floatfix]{revtex4-1}
\usepackage[utf8x]{inputenc}
\usepackage[english]{babel}
\usepackage[T1]{fontenc}
\usepackage{lmodern}
\usepackage{amssymb}
\usepackage{bbold}
\usepackage{amsmath}
\usepackage{amsthm}
\usepackage[pdftex]{graphicx}
\usepackage{epstopdf}
\usepackage{cancel}
\usepackage{lipsum}
\usepackage{dcolumn}
\usepackage{bm}
\usepackage{color}
\usepackage{relsize,dsfont,mathrsfs,empheq,verbatim,upgreek}
\usepackage{etoolbox}
\usepackage[caption = false]{subfig}
\usepackage{wrapfig}
\usepackage{datetime}
\usepackage{pgfplots}
\usepackage{times}
\usepackage{hyperref}
\hypersetup{%
   bookmarksopen=false,
   bookmarksnumbered=true,
   pdfnewwindow=true,
   unicode=false,
   colorlinks=true,%
   citecolor=blue,
   linkcolor=black,
   urlcolor=blue,
   filecolor=blue
   }  
\definecolor{darkblue}{rgb}{0,0,0.5}
\definecolor{grey25}{rgb}{0.25,0.25,0.25}
\definecolor{DarkOrange1}{rgb}{1.00,0.50,0.00}
\definecolor{lightblue}{rgb}{0,0,0.8}
\definecolor{royalblue}{rgb}{0.25,0.41,0.88}
\definecolor{lightcyan}{rgb}{0,0.8,0.8}
\definecolor{MediumVioletRed}{rgb}{0.78,0.08,0.52}
\definecolor{gold}{rgb}{1.00,0.84,0.00}
\definecolor{OrangeRed}{rgb}{1.00,0.27,0.00}
\definecolor{darkgreen}{rgb}{0,0.5,0}
\definecolor{viola}{rgb}{0.58,0,0.72}
\definecolor{ocra}{rgb}{0.673,0.276,0.05}
\definecolor{turchesevioletto}{rgb}{0.354,0.17,0.48}

\DeclareMathOperator{\Tr}{Tr}

\begin{document}
\newcommand{\freiburg}{Physikalisches Institut, Albert-Ludwigs-Universität Freiburg, Hermann-Herder-Stra{\ss}e 3, D-79104 Freiburg i.~Br., Federal Republic of Germany}
\newcommand{\frias}{Freiburg Institute for Advanced Studies, Albert-Ludwigs-Universit\"at Freiburg, Albertstr.~19, D-79104 Freiburg i.~Br., Federal Republic of Germany}
\newcommand{\parma}{Dipartimento di Scienze Matematiche, Fisiche e Informatiche, Universit\`a di Parma, Campus Universitario, Parco area delle Scienze n. 7/a, 43124 Parma, Italy}
\newcommand{\infn}{INFN, Sezione di Milano Bicocca, Gruppo Collegato di Parma, 43124 Parma, Italy}
\newcommand{\usal}{Departamento de F\'isica Fundamental, Universidad de Salamanca, E-37008 Salamanca, Spain}

\title{Quantum transport between finite reservoirs}%
\author{Giulio Amato}
\affiliation{\freiburg}
\affiliation{\parma}
\affiliation{\infn}
\author{Heinz-Peter Breuer}
\affiliation{\freiburg}
\affiliation{\frias}
\author{Sandro Wimberger}
\affiliation{\parma}
\affiliation{\infn}
\author{Alberto Rodríguez}
\affiliation{\freiburg}
\affiliation{\usal}
\author{Andreas Buchleitner}
\email[]{a.buchleitner@physik.uni-freiburg.de}
\affiliation{\freiburg}
\affiliation{\frias}
\begin{abstract}
When driven by a potential bias between two finite reservoirs, the particle current across a quantum system evolves from an initial loading through a coherent, followed by a metastable phase, and ultimately fades away upon equilibration. We formulate a theory which fully accounts for the associated, distinct time scales, and identifies the parameter dependence of the decay rate which ultimately controls the convergence towards equilibrium. Our formalism guarantees total particle number conservation and fundamental consistency between macroscopic and internal currents flowing in the system. We furthermore establish a clear imprint of the fermionic or bosonic particle character on the resulting conductance.
\end{abstract}
\maketitle

The field of quantum transport theory, originally developed in the realm of electronic systems \cite{Datta1995,Dit1998,Ram1998,Beenakker1991,Datta2005}, finds an ideal experimental platform in ultracold atoms, where already established as well as new observations have been reported \cite{Bra2012,Schneider2012,Stadler2012,Gat2012,Bra2013,Labouvie2015,Krinner2013,
Krinner2015,Husmann2015,Krinner2016a,Hausler2017,Krinner2017,Leb2018,Lebrat2019}. Ultracold atom setups allow for a fine parametric control, e.g. tuning the interparticle interactions, provide good isolation from unwanted degrees of freedom, offer the possibility to study transport with bosonic or fermionic carriers, and enable the observation of bona fide many-particle interference phenomena. Interestingly, recent ultracold atom transport experiments \cite{Krinner2017,Leb2018,Lebrat2019} have witnessed the intrinsically finite, and hence non-stationary, nature of particle reservoirs, and the related dynamical consequences, such as non-stationary, decaying currents, and the relaxation into a final equilibrium. 

The dynamics of the degrees of freedom connecting two reservoirs in a transport setup has been extensively studied within the framework of open quantum systems \cite{HaEs2006,EsHa2007,Wic2007,Pep2010,Pro2012,Zni2013,Asa2013,Ber2013,Iva2013,Kul2014,Kul2015,Kul2016,Seg2016,Hofer2017,Kol2017,Kol2018,Kolovsky2019}, mostly under the assumption of stationary environment, although attempts to include the time evolution of the reservoirs have been put forth \cite{Ponomarev2006,Bruderer2012,Schaller2014,Gallego-Marcos2014,Purk2017,deVega2018}. Nevertheless, to the best of our knowledge, a formalism able to describe the non-trivial concurrent quantum dynamics of  the system, together with the classical dynamics of the reservoirs, while capturing all the emerging different dynamical regimes, is lacking.

It is the purpose of this work to present a master equation approach which, under the assumption of permanently thermalized reservoirs and the premise of particle number conservation in the entire composite system (as in the experimental realizations with cold atoms \cite{Krinner2017,Leb2018}), allows for the description of the coupled evolution of system and environment, captures all distinct emerging time scales, and accounts for the potential impact of many-particle interference upon quantum transport.

\begin{figure}
\centering
\includegraphics[width=0.95\columnwidth]{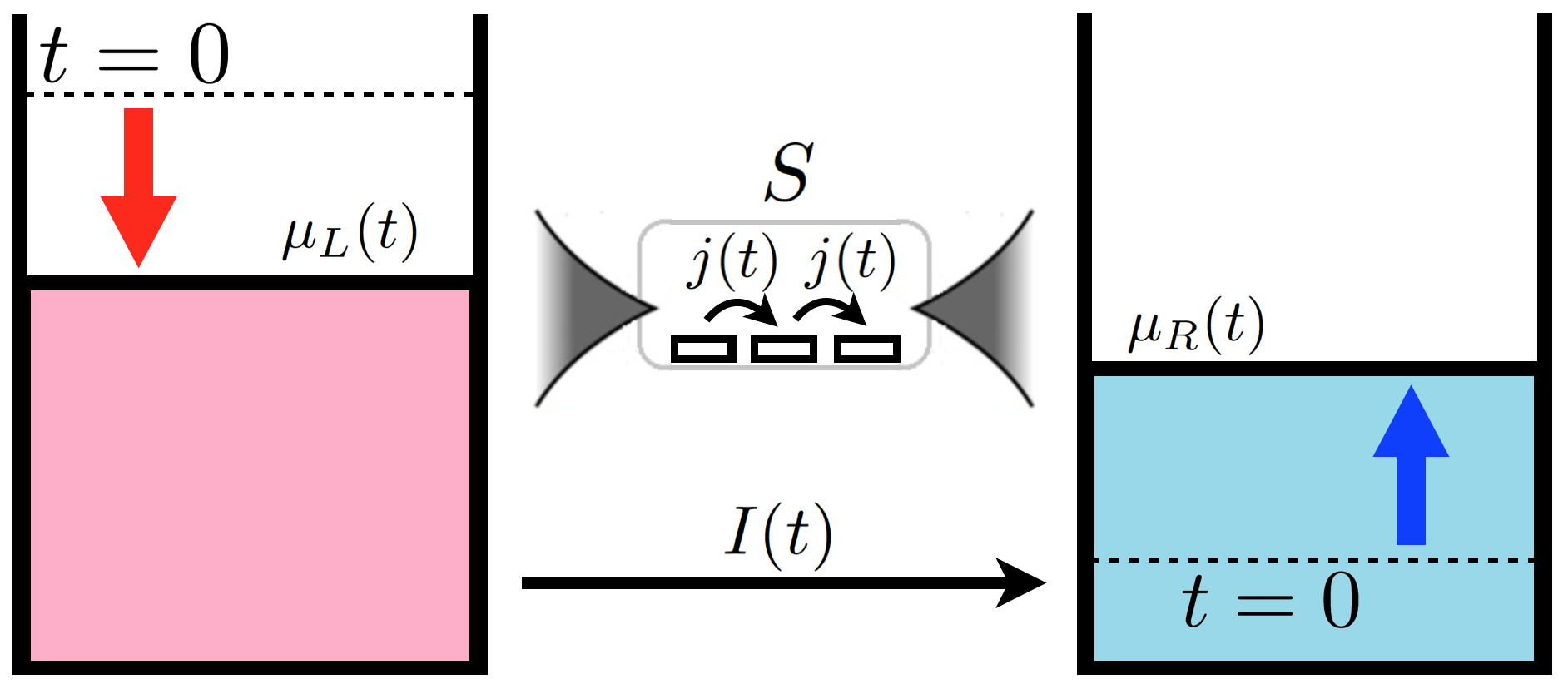}
\caption{Schematic representation of the evolution of the population distribution of a finite number of particles over two reservoirs and a lattice connecting them: A finite population difference between the reservoirs at $t=0$, corresponding to an imbalance of the associated chemical potentials $\mu_{L,R}$, drives microscopic currents $j(t)$ across the initially empty lattice. These generate a macroscopic current $I(t)$ which reduces the reservoir bias. The current ceases upon equilibration between both reservoirs.}
\label{fig:schemeEquilibration}
\end{figure}

We consider an open system $S$ made up of $M$ lattice sites, which is connected, via its first and last site, respectively, to two finite (bosonic or fermionic) reservoirs, left ($L$) and right ($R$) (see Fig.~\ref{fig:schemeEquilibration}). The non-interacting system Hamiltonian reads 
$H_S = \varepsilon_S \sum_i a_i^{\dagger} a_i - J \sum_i (a_i^{\dagger} a_{i+1}+a_{i+1}^{\dagger}a_i)$,
where $ \varepsilon_S $ is the uniform on-site energy and $ J $ denotes the nearest neighbour tunnelling strength. 
Given earlier results on the stationary case \cite{Amato2019}, together with insights from the use of time-dependent projectors to 
account for the non-stationarity of the reservoir, we conclude that the open system evolution is generated by the master 
equation
\begin{equation}
\begin{aligned}
 \dot{\rho}_{S} (t) = &
 - i [ H_S , \rho_S (t) ] \\
  & +  \gamma_L^+ (t) \mathcal{D} [ {a}_1^{\dagger} ] [ {\rho}_S (t) ] + \gamma_L^- (t) \mathcal{D} [ {a}_1 ] [ {\rho}_S (t) ]  \\
  & +  \gamma_R^+ (t) \mathcal{D} [ {a}_M^{\dagger} ] [ {\rho}_S (t) ] + \gamma_R^- (t) \mathcal{D} [ {a}_M ] [ {\rho}_S (t) ],
 \end{aligned}
\label{eq:meLindbladtime}
\end{equation}
with the dissipator 
$\mathcal{D}[{A}][\rho_S] = {A} \rho_S  {A}^{\dagger}-\frac{1}{2}\{{A}^{\dagger}{A},\rho_S\}$,
where the ---a priori unknown--- rates $\gamma_{L,R}^+(t)$ [$\gamma_{L,R}^- (t)$] account for the particle injection (extraction) into (from) the left and right reservoir at the first and last lattice site, respectively, with an explicit time dependence which is inherited from the non-trivial evolution of the reservoir states.
Note that the spatially localized dissipation at sites $1$ and $M$ seeds a competition between coherent system dynamics and decoherent effects induced by the reservoirs \cite{Note1}.

Assuming that the characteristic time scale of the reservoir degrees of freedom is much shorter than that of the dynamical evolution in $S$, we describe the reservoirs by 
grand canonical thermal states with a fixed common temperature and time-dependent chemical potentials, ${\varrho}_{L,R} (t) \propto \exp \{-\beta[ {H}_{L,R} - \mu_{L,R} (t) \hat{N}_{L,R}]\}$, 
where the Hamiltonians $H_{L,R}$ define the thermally occupied reservoir modes and $\hat{N}_{L,R}$ are the corresponding total particle number operators.   

To ensure the conservation of the total particle number distributed over the lattice and both reservoirs, we require that the dissipative gain/loss of particles on the first (last) system site be precisely balanced by the change of particle number in the left (right) reservoir, i.e., 
\begin{equation}
\dot{n}_{1,M} (t ) \big|_\textrm{diss} = - \dot{N}_{L,R} (t).
\label{eq:consPartic}
\end{equation}
Since $n_j(t)=\Tr_S[a_j^\dagger a_j \rho_S(t)]$, it follows from Eq.~\eqref{eq:meLindbladtime} that
\begin{equation}
\dot{n}_{1,M} (t ) \big|_\textrm{diss} = -  \left[\gamma_{L,R}^- (t) \mp \gamma_{L,R}^+ (t) \right] n_{1,M} (t ) + \gamma_{L,R}^+ (t), 
\label{eq:dissn1}
\end{equation}
where the upper (lower) sign stands for bosons (fermions). 
On the other hand, we take the dynamics of the reservoir particle number to be governed by the 
 classical rate equation
\begin{equation}
  \dot{N}_{L,R} (t) = \gamma \left[ n_{1,M} (t )- n_{L,R} (\varepsilon_{S}, t)\right],
\label{eq:NL}
\end{equation}
which states that the rate of change of the reservoir population $N_{L,R}$ is proportional to the population offset between the first (last) lattice site and the resonant reservoir energy level at energy $\varepsilon_S$. 
Subdominant coupling to non-resonant reservoir states is neglected here, consistent with a Markov approximation which assumes perturbative coupling strengths between system and environment. The latter together with the reservoir's density of states determine the time independent coupling constant $\gamma$ \cite{Hofer2017,Amato2019} which we assume to be the same for both reservoirs, 
to simplify our subsequent formulae. (We present the general case $\gamma_L\neq\gamma_R$ in Ref.~\cite{Amato2019}.)

Matching Eqs.~\eqref{eq:dissn1} and \eqref{eq:NL} by Eq.~\eqref{eq:consPartic} determines the time dependence of the rates,
\begin{subequations}
\begin{eqnarray}
\gamma_L^+ (t) &=& \gamma n_L (\varepsilon_{S}, t ), \\
\gamma_L^- (t) &=& \gamma [ 1\pm n_L (\varepsilon_{S}, t)],
\end{eqnarray}
\label{eq:tDepRates}
\end{subequations}
and equivalently for $\gamma_R^{\pm}(t)$.
Note that the time dependent rates are positive for all times, which guarantees a well defined evolution for every initial state $\rho_S$ in Eq.~\eqref{eq:meLindbladtime} \cite{Gorini1976,Lindblad1976a} and Markovian dynamics of the system degrees of freedom \cite{BrLa2016}. Moreover, the expressions above are a straightforward time dependent generalization of the rates which result for an analogous master equation treatment derived for stationary reservoirs \cite{Hofer2017}.

The master equation \eqref{eq:meLindbladtime}, with time-dependent rates \eqref{eq:tDepRates}, and the classical rate equations \eqref{eq:NL} for the reservoir populations, provide a description of the system-environment concurrent time evolution that explicitly ensures total particle number conservation. Note that the state of the reservoirs is determined by the chemical potentials via, e.g., $n_L (\varepsilon, t)=\left(e^{\beta[\varepsilon-\mu_L(t)]}\mp 1 \right)^{-1}$ and 
$N_L(t)=\int_{E_0}^\infty d\varepsilon\,D(\varepsilon)\,n_L(\varepsilon,t)$, where $D(\varepsilon)$ is the reservoir density of states and $E_0$ its lowest energy state. Hence, the system-reservoir dynamics are governed by a set of coupled \emph{nonlinear} differential equations. This mathematical framework constitutes the first significant result of this letter.

We proceed to investigate the system dynamics via the single-particle density matrix (SPDM), $\sigma_{jk} (t) = \Tr_S [ a_j^{\dagger} a_k \rho_S (t)]$, which provides access to on-site populations $ n_l (t ) = \sigma_{ll} (t) $, local site-to-site currents $ j_{l,l+1} (t) \equiv i J [ \sigma_{l+1,l} (t) -\sigma_{l,l+1} (t)]$ and long range coherences, for $|j-k|>1$. To ease our treatment, it is  convenient to recast the differential equations for the reservoirs [Eq.~\eqref{eq:NL}] in terms of the chemical potentials \cite{Amato2019}. Then, the resulting coupled evolution equations form a closed set in the variables $\{\sigma_{jk}(t),\mu_L(t),\mu_R(t)\}$. 
The reservoirs are modelled as identical 3D anisotropic harmonic traps,
resembling the setup of recent transport experiments with ultracold atoms \cite{Krinner2017,Leb2018}, and allowing for analytical calculations \cite{Amato2019} and fast numerical solution
of the coupled nonlinear dynamical equations.

\begin{figure}
\centering
\includegraphics[width=.9\columnwidth]{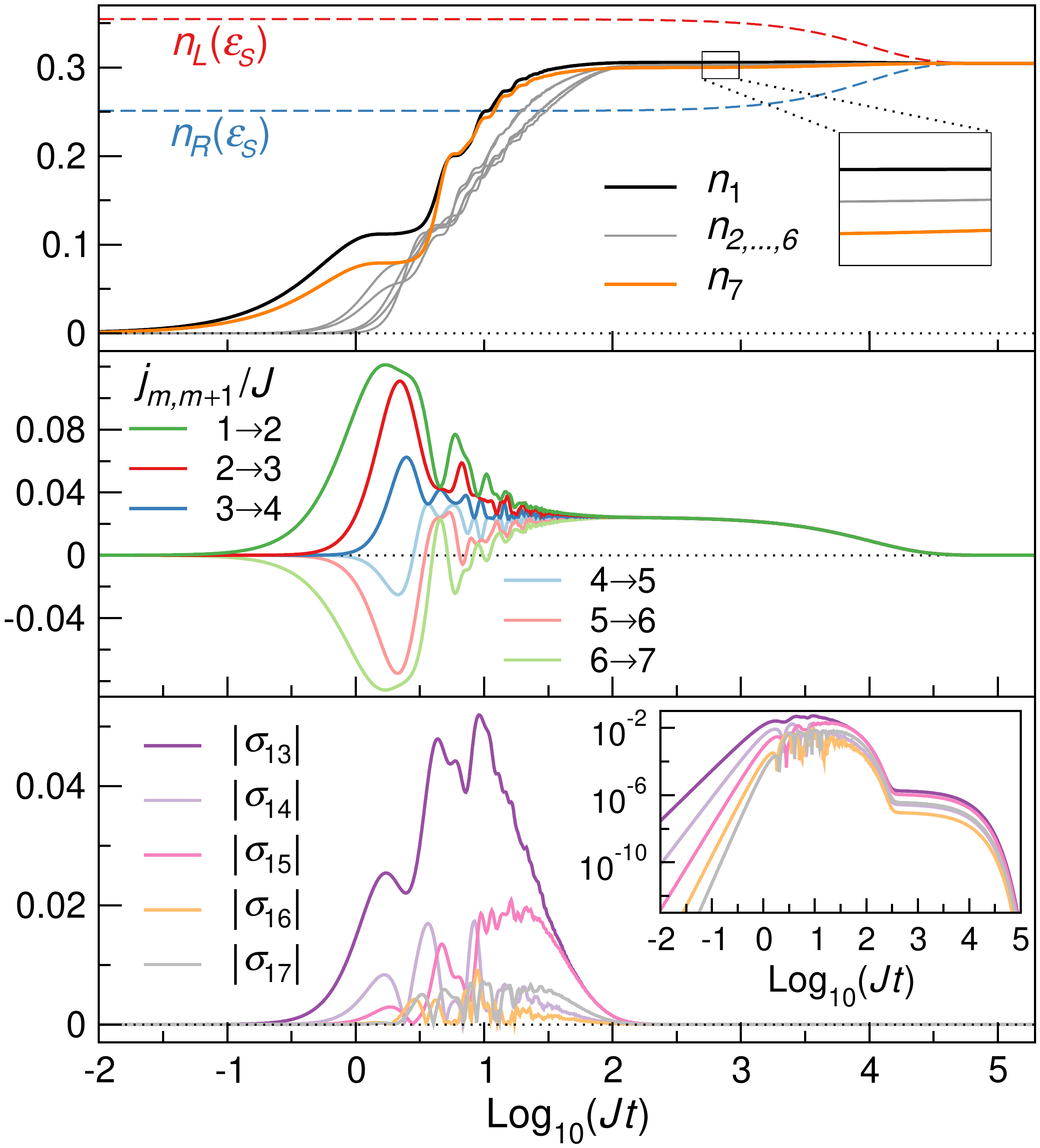}
\caption{Fermionic transport between non-stationary reservoirs. The panels show the evolution of (top) lattice site populations $n_l(t)$ and reservoir resonant occupations $n_{L,R}(\varepsilon_S,t)$, (middle) site-to-site currents $j_{l,l+1}(t)$, and (bottom) long-range coherences $\{\sigma_{1j}\}_{j=3,\ldots,7}$. 
The parameters used are $M=7$, $\varepsilon_S=2J$, $\gamma=0.5J$, $\beta J=1$, $\omega_x=\omega_y=0.2J$, $\omega_z=0.05J$, $\mu_L(0)=1.401J$, $\mu_R(0)=0.907J$, corresponding to $N_L(0)= 1500$, $N_R(0)=1000$, $n_L(\varepsilon_S,0)=0.355$ and $n_R(\varepsilon_S,0)=0.251$. 
The final equilibrium condition is characterized by $\mu^\infty=1.174$, $n^\infty=0.305$, $N^\infty\simeq1249$ [see discussion after Eq.~\eqref{eq:Currentconsistency}].
}
\label{fig:TransFerFinRes}
\end{figure}
In Fig.~\ref{fig:TransFerFinRes}, we show an overview of the full time evolution of various single particle quantifiers of quantum transport between nonstationary reservoirs, where the lattice is initially empty, and the conserved total particle number is $N_0=N_L(0)+N_R(0)\gg 1$. The short time dynamics exhibits coherent oscillations on the lattice, progressively affected by dissipation due to the interaction with the baths, followed by a (quasi)stationary state which mediates the redistribution of particles between the reservoirs and eventually leads to a final equilibrium with vanishing net current.

The very early stage of the dynamics (particle loading into the lattice) is governed by a power-law growth of all SPDM elements, $|\sigma_{jk}(t \rightarrow 0)| \propto (J t)^{ M- |j+k -(M+1)|}$ for $Jt\lesssim \gamma^{-1}$ (see bottom inset in Fig.~\ref{fig:TransFerFinRes}), as derived from an iterative solution of the equations around $t=0$ \cite{Amato2019}.
Afterwards, single particle interferences give rise to characteristic fluctuations of the observables, while long-range coherences reach their maximum values. We find that the damping of this coherent regime, induced by the dissipative coupling to the reservoirs, is controlled by a time scale $\tau_{\textrm{rel}}\propto M^3/\gamma$ \cite{Bi2014,Med2014,Zni2015,Kol2018}, which can be inferred from the description of dissipation-induced decoherence via an effective non-hermitian Hamiltonian \cite{Gardiner1985,See2017,Amato2019}.

The relaxation of single-particle interference yields a slowly varying metastable state, in which the SPDM long-range coherences are strongly suppressed and the time variation of observables, such as onsite populations and site-to-site currents, is negligible compared to their instant expectation values. Using the latter approximation in the equations of motion leads to the homogenization of all site-to-site currents, 
\begin{equation}
 j_{l,l+1}(t)\approx j (t ) = \frac{ 2 \gamma J^2 }{ 4 J^2 + \gamma^2 } \Delta n (t),
\label{eq:MetastableCurrent}
\end{equation}
and a ladder like structure for the lattice populations (see upper inset in Fig.~\ref{fig:TransFerFinRes}), 
\begin{subequations}
\begin{eqnarray}
n_{1,M} (t) &\approx& \bar{n} (t ) \pm \frac{ \gamma^2 }{ 2(4 J^2 + \gamma^2) } \Delta n (t ) , \\
n_{m } (t ) &\approx&  \bar{n} (t ), \qquad 2\leqslant m\leqslant M-1,
\end{eqnarray}
\label{eq:MetastablePopulations}
\end{subequations}
given in terms of
\begin{subequations}
\begin{eqnarray}
  \Delta n (t) &=&  n_L ( \varepsilon_S ,t ) - n_R ( \varepsilon_S , t), \label{eq:deltan}\\
  \bar{n} (t) &=&  \left[ n_L ( \varepsilon_S ,t ) + n_R ( \varepsilon_S ,t )\right]/2.
  \label{eq:barn}
\end{eqnarray}
\label{eq:deltanbarn}
\end{subequations}
Equations \eqref{eq:MetastableCurrent} and \eqref{eq:MetastablePopulations} are a time dependent generalization of the non-equilibrium steady state that emerges in the case of stationary reservoirs \cite{Asa2013}. Hence, the metastable regime can be seen as a succession of non-equilibrium (quasi)stationary states, driven by time evolving reservoirs. 

The metastable dynamics effectively enable a direct particle exchange between the reservoirs and the emergence of a macroscopic current $I(t)$, defined by 
\cite{Datta1995,Lan1957,But1986,Mei1992,Krinner2017,Leb2018}
\begin{equation}
I (t ) := - \frac{1}{2} \frac{d}{dt } \Delta N (t) ,
\label{eq:bigCurrent}
\end{equation}
where $ \Delta N (t) = N_L (t) - N_R (t)$. From Eqs.~\eqref{eq:NL}, \eqref{eq:MetastableCurrent} and \eqref{eq:MetastablePopulations}, one can verify that 
\begin{equation}
  I (t ) = j(t),
\label{eq:Currentconsistency}
\end{equation}
i.e., the internal and macroscopic currents coincide.
This fundamental consistency of the transport process follows from the equations of motion and is always 
guaranteed by our formalism after the onset of metastability.

As $t\to\infty$, the reservoir-system dynamics converge to the final equilibrium state, characterized by vanishing currents and a uniform population $n^\infty=\left[e^{\beta(\varepsilon_S-\mu^\infty)}\mp 1\right]^{-1}$ of all resonant energy levels. The reservoir equilibrium chemical potential $\mu^\infty$ can be numerically found from the particle number conservation condition, $N_0=2N^\infty + M n^\infty$, where $N^\infty=N_{L,R}(t\to\infty)$.

\begin{figure}
\centering
\includegraphics[width=\columnwidth]{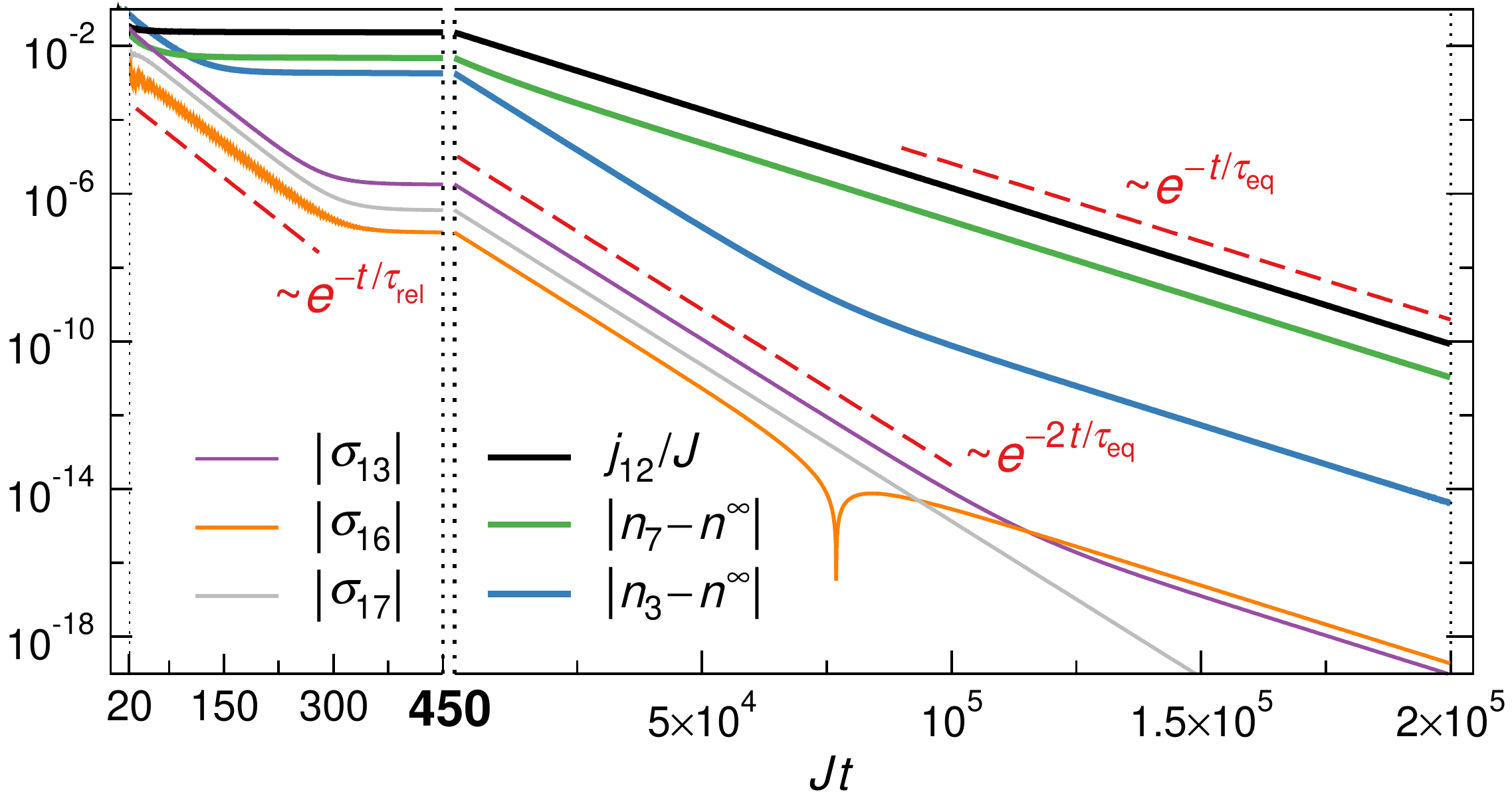}
\caption{Long time evolution of local current $j_{12}$ (thick black line), populations $n_3$ and $n_7$ (thick colored lines),  and long-range coherences $\{\sigma_{1j}\}_{j=3,6,7}$ (thin colored lines) towards their respective equilibrium values. 
Note the different scales on the time axis, concatenated at the value highlighted in bold. 
Dashed red lines indicate the exponential decay on time scales $J\tau_{\textrm{rel}}=27.833$, obtained numerically from an effective non-hermitian Hamiltonian \cite{Amato2019}, and $J\tau_\textrm{eq}= 1.0265\times 10^{4}$, obtained from Eq.~\eqref{eq:taueq}. Same model parameters as used in Fig.~\ref{fig:TransFerFinRes}.}
\label{fig:AsympDyn}
\end{figure}

In order to fully understand the time dependence of observables in the metastable regime, and hence their approach towards equilibrium, we need to infer the asymptotic dynamics of $\Delta n(t)$ and $\bar{n}(t)$ [Eqs.~\eqref{eq:deltanbarn}]. Manipulation of the reservoir equations of motion [e.g., Eq.~\eqref{eq:NL}] in leading order around the equilibrium condition, in combination with the metastable form  \eqref{eq:MetastablePopulations} of the on-site populations yields \cite{Amato2019}
\begin{subequations}
\begin{eqnarray}
 \Delta n(t\to\infty) & \propto & e^{-t/\tau_\textrm{eq}}, \\
 |\bar{n}(t\to\infty)-n^\infty| &\propto& e^{-2 t/\tau_\textrm{eq}},
\end{eqnarray}
\label{eq:deltanbarnExpt}
\end{subequations}
where the dependence of the emergent equilibration time scale $\tau_{\textrm{eq}}$ on the different physical parameters of the system and reservoirs is analytically obtained \cite{Amato2019}, and in the case $N_0\gg M$ can be approximated as 
\begin{equation}
 \tau_\textrm{eq} \approx \frac{\Delta N (0)}{\Delta n (0)} \frac{4 J^2 + \gamma^2}{4 \gamma J^2},
 \label{eq:taueq}
\end{equation}
with no dependence on $M$.
The exponential forms \eqref{eq:deltanbarnExpt} carry over to $j(t)$, $I(t)$, $\Delta N(t)$, $\Delta\mu(t)\equiv \mu_L(t)-\mu_R(t)$, on-site populations and long-range coherences, and are confirmed by numerical simulations, as shown in Fig.~\ref{fig:AsympDyn}. Furthermore, it follows that  
\begin{equation}
 I(t)=\frac{1}{2\tau_\textrm{eq}}\Delta N(t) ,
\end{equation}
i.e., our formalism predicts an exponentially decaying, ohmic macroscopic current between equilibrating reservoirs, independently of 
the bosonic or fermionic nature of the particles, in qualitative agreement with cold-atom transport experiments \cite{Krinner2017}. 

\begin{figure}
\centering
\includegraphics[width=.9\columnwidth]{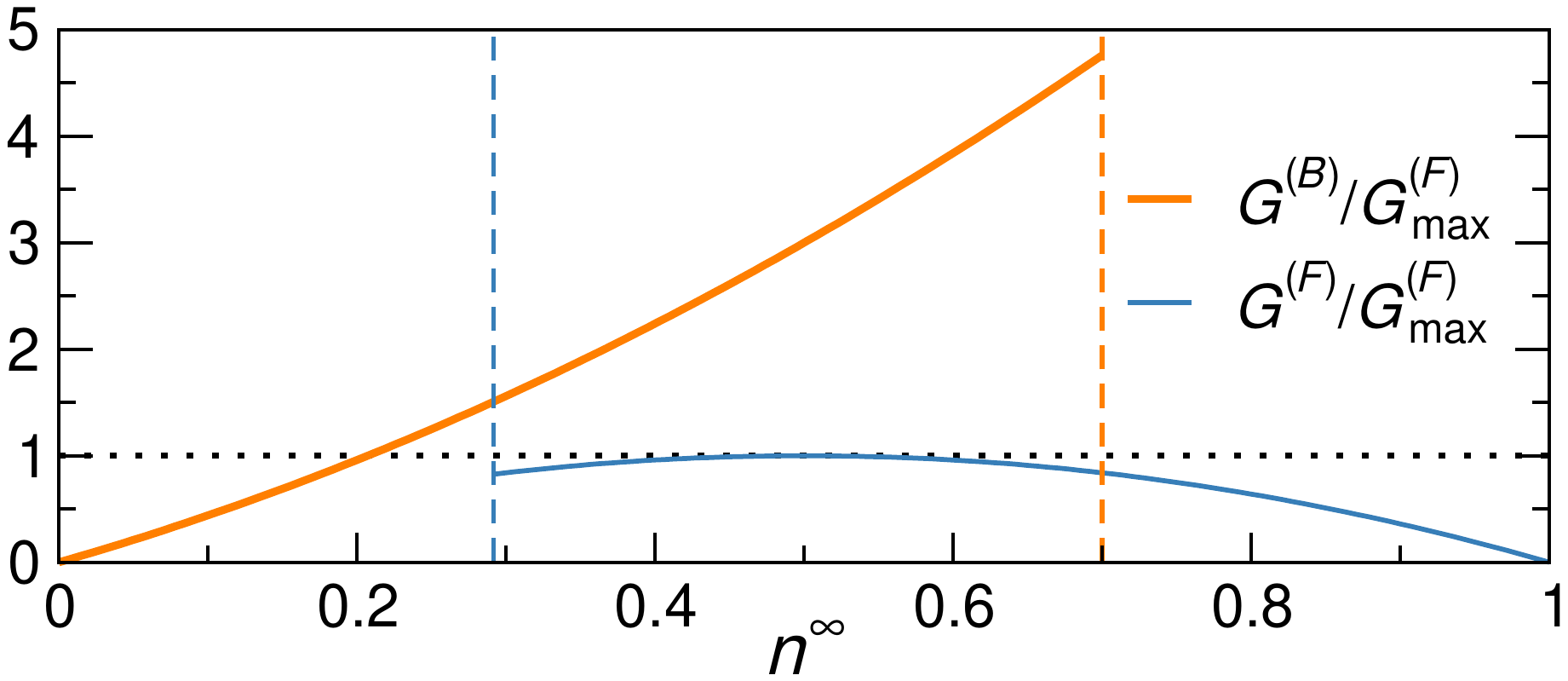}
\caption{Conductance $ G $ of a fermionic (thin line) and of a bosonic (thick line) channel, as a function of the equilibrium population $n^\infty$. The horizontal dotted line indicates the upper bound $G^{(F)}_\textrm{max}$ for fermionic conductance.
Vertical dashed lines illustrate the allowed lower (upper) limit on $n^\infty$ for fermions (bosons) with $\beta J=0.5$, $\varepsilon_S=2J$, $\omega_x=\omega_y=0.2J$, $\omega_z=0.05J$.}
\label{fig:Conductance}
\end{figure}

Our treatment also provides an expression for the conductance $ G $ of the lattice. From the known exponential behaviour of $I(t)$ [$j(t)$] and $\Delta\mu(t)$ in the metastable regime, we find 
\begin{equation}
 G := \frac{I(t)}{\Delta\mu(t)} =\beta n^\infty(1\pm n^\infty) \frac{2\gamma J^2}{4J^2+\gamma^2},
 \label{eq:conductanceG}
\end{equation}
where the upper (lower) sign stands for bosons (fermions). 
This expression immediately reveals that the conductance $G^{(F)}$ of a fermionic channel is fundamentally bounded from above, $G^{(F)}\leqslant G^{(F)}_\textrm{max} \equiv\beta \gamma J^2 /2 (4J^2+\gamma^2)$ as shown in Fig.~\ref{fig:Conductance}, where $G^{(F)} $ is plotted versus   $n^\infty$ ---which is mainly controlled by the total particle number $N_0$, the reservoir temperature and the value of the resonant energy $\varepsilon_S$. 
In contrast, the conductance of a bosonic channel increases monotonically with $n^\infty$, and can be several times larger than the fermionic conductance ---even in transport configurations with comparable equilibrium occupation number $ n^{\infty} $. Note that, since $\mu^\infty$ is bounded from below (above) by the lowest reservoir energy level $E_0$ for fermions (bosons), the asymptotic occupation $n^\infty$ has the temperature dependent lower (upper) limit $\left(e^{\beta[\varepsilon_S-E_0]}\mp 1 \right)^{-1}$.

The dependence of the conductance on the quantum statistics of the particles can be understood by noticing that $n^\infty(1\pm n^\infty) = \lim_{t\to\infty} (\langle \hat{n}_j^2\rangle -\langle \hat{n}_j\rangle^2) $ \cite{Note2} 
on the right hand side of Eq.~\eqref{eq:conductanceG}, hence $ G $ is directly proportional to the variance of the equilibrium population. The latter involving the expectation value of a two-particle observable, makes it sensitive to many-particle interference \cite{Tichy2010,Mayer2011,Bru2018}, which in turn depends on the bosonic or fermionic nature of the particles. 
Furthermore, our formalism also permits to access the dynamics of higher-order observables, such as the two-particle density matrix, and to witness fingerprints of the many-particle nature of the transport process considered, e.g., in distinct fluctuation properties of the internal metastable current for fermionic and bosonic particles \cite{Amato2019}.

We have derived a set of quantum-classical master equations that describes the coupled dynamics of a system and two finite size (bosonic or fermionic) reservoirs, which evolve through grand canonical thermal states with time-dependent chemical potentials. This approach relies on local coupling between system and reservoirs and on the validity of the Markovian approximation. 
Our formalism guarantees particle conservation and fundamental consistency in the transport regime between macroscopic and internal currents, which exhibit a generic exponential decay 
 (in qualitative agreement with experimental observations \cite{Krinner2017}) characterized by an equilibration time scale which has been analytically obtained.
Furthermore, our approach entails that the system conductance is sensitive to the fermionic or bosonic nature of the carriers, highlighting the significance of 
many-particle interference effects in quantum transport through open systems. 

\begin{acknowledgments}
The authors are grateful to the German Research Foundation (DFG project WI3426/7-1), to the state of Baden-W\"urttemberg through bwHPC, and thank Martin Lebrat for helpful discussions.
GA furthermore acknowledges support by Fondazione Grazioli and by DAAD.
\end{acknowledgments}
%
\end{document}